\documentclass[amsmath,amssymb,aps,floats,prl,twocolumn,showpacs,unsortedaddress,superscriptaddress]{revtex4}

\usepackage{graphicx}
\usepackage{amssymb}
\usepackage{amsmath}
\usepackage{epstopdf}
\usepackage{color}
\usepackage[normalem]{ulem}

\newcommand{\ie}{\emph{i.e.} }

\newcommand{\be}{\begin{eqnarray}}
\newcommand{\ee}{\end{eqnarray}}
\newcommand{\bfig}{\begin{figure}}
\newcommand{\efig}{\end{figure}}

\newcommand{\black}{\color{black}}

\newcommand{\mnfesi} {Mn$_{0.9}$Fe$_{0.1}$Si}

\renewcommand{\arraystretch}{1.25} 

\begin{document}

\title{ Multiple magnetic states within the $A$-phase determined by field-orientation dependence of \mnfesi}

\author{Peter E. Siegfried }
\affiliation{Department of Physics, University of Colorado, Boulder, CO 80309, USA}%
\author{Alexander C. Bornstein$^{\dagger}$ }
\affiliation{Department of Physics, University of Colorado, Boulder, CO 80309, USA}%
\author{Andrew C. Treglia}
\affiliation{Department of Physics, University of Colorado, Boulder, CO 80309, USA}%
\author{Thomas Wolf}
\affiliation{Institute for Solid-State Physics, Karlsruhe Institute of Technology, D-76021 Karlsruhe, Karlsruhe, Germany}%
\author{Minhyea Lee}
\email{minhyea.lee@colorado.edu}
\affiliation{Department of Physics, University of Colorado, Boulder, CO 80309, USA}%

\date{\today}

\begin{abstract}

We report three distinct regions within the $A$-phase in Fe-doped MnSi, based on the evolution of magnetoresistance and the Hall effect as a function of orientation of applied field. 
Fe impurities as pinning centers and crystalline anisotropy are found non-negligible only at the boundary of the $A$-phase. Electrical transport characteristics unique to the $A$-phase not only remain robust, but also indicate a freely rotating skyrmion lattice, decoupled from underlying crystal structure or impurity pinning.

\end{abstract}
 
\maketitle

{\emph {Introduction }} 
Cubic B20 structured intermetallic magnets provide a rare opportunity to study unique properties of  the non-trivial magnetic texture known as the magnetic skyrmion lattice (SkL) as a collection of particle-like individual skyrmions. 
A small pocket in the $H$-$T$ phase diagram, where the SkL emerges is called the $A$-phase  \cite{Thessieu1997,Muhlbauer2009}.

The similarities between SkL and vortex lattice phases in type-2 superconductors has been recognized and many efforts were devoted to exploring possible skyrmion states~\cite{Wilhelm2011, Janson2014, Lobanova2016}, analogous to those exhibited by superconducting vortices. In particular, the presence of quenched disorders in superconducting states have been known to drive the system into a variety of vortex states that exhibit unique electrical transport characteristics and a rich phase diagram~\cite{Blatter1994}. 
While the role of quenched disorders in the SkL has been considered theoretically in a dynamic limit ~\cite{Reichhardt2015}, the experimental examination is scarce.
Quenched disorders are expected to cause deformation of the SkL arrangement~\cite{Chapman2013} and to influence the strength of the emergent field and scattering rates of the conduction electrons. Therefore, tracking 
the electrical transport properties will shed light on the nature of the $A$-phase and possible discernible states within.

Within the $A$-phase, the conspicuous topological Hall effect is observed, which arises from the emergent gauge field $\mathbf{b}$ expressed as, $b_i = \frac{\Phi_0}{8\pi}\epsilon_{ijk}\hat n\cdot(\partial_j \hat n \times \partial_k \hat n),$ where $\epsilon_{ijk}$ is the Levi-Civita symbol with the indices running over $x,y,$ and $z$, $\hat n(\mathbf{r})$ is the unit vector of the magnetization $\mathbf M(\mathbf{r})$, and $\Phi_0 = h/|e|$~\cite{Bogdanov1989,Rossler2006,Schulz2012}. The magnitude of this gauge field is estimated to as large as 20 -- 45 T for Fe-doped MnSi, giving rise to a highly unusual $H$-profile of the Hall signal ~\cite{Chapman2013}.

In this paper, we report the field angle dependence of the Hall signal, $\rho_{yx}$ and magnetoresistance (MR) in the presence of quenched disorders viz. Fe-doping in MnSi. 
We focus on near and inside of the $A$-phase in the phase diagram to reveal discernible areas within the $A$-phase  whose origin could be attributed to weak crystalline anisotropy and quenched disorders. 
We find that the $A$-phase is divided into three distinct regions, which we call Region I, II and III : 
Region I shows the most prominent  crystalline anisotropy and pinning effects, yet overall, both  turn out relatively weak. 
Region II corresponds to the center of the $A$-phase and is characterized by the angle-{\emph{independent}} MR, occurring at the same  magnitude of field, $H_m$ at which the THE has a maximum magnitude at a given $T$. 
Region III is found at elevated $T$ and $H$ above Region II, where the $\rho_{yx}(H)$ remains constant with increasing $H$ and both MR and Hall signals as a function of the field angle recover the usual functional forms, indicating the background magnetization smoothly increases to the full spin-polarized state.

\begin{figure}[ht]
\begin{center}
\includegraphics[width=0.8\linewidth]{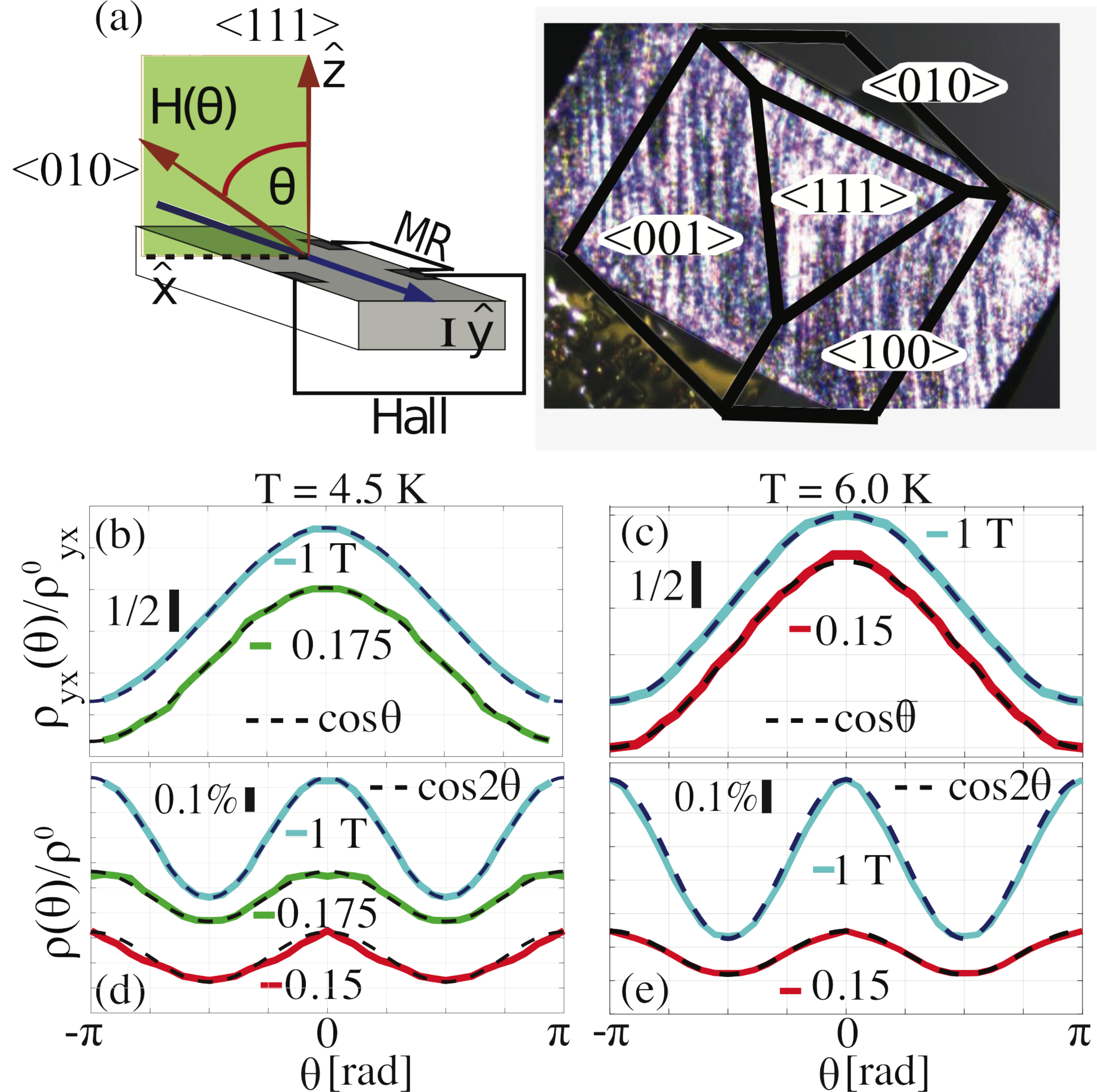}
\caption{\small (a)(Left) Measurement configuration with a bar-shaped \mnfesi. $\theta$ measured from the $\langle111\rangle$ direction sweeps through $\langle010\rangle$.  (Right)  Image of \mnfesi~crystal with crystallographic faces labeled via X-ray single crystal diffraction. (b) Normalized Hall resistivity $\rho_{yx}(\theta)/\rho_{yx}^0$ at fixed $H = 0.15$ (red), 0.175 (green) and 1 T (light blue) at $T=4.5$ K and (c) at 6 K. (d) $\theta$ dependence of normalized MR at $T=4.5$ K and (e) at 6 K. 
Broken lines establish $\cos\theta$ and $\cos2\theta$ for Hall and MR respectively.
}
\label{AngDep}
\end{center}
\end{figure}

{\emph {Methods }} 
Single crystals of \mnfesi~($T_C = 6.9$ K) used were grown by the Bridgman technique and cut into rectangular prisms for a standard 6-contacts~geometry.
Samples were mounted on a home-built rotation probe which allows full $2\pi$ rotation of the applied magnetic field in the $xz$-plane with the current $\mathbf{I} \parallel \hat y$, as shown in Fig.~\ref{AngDep}(a). Polar angle $\theta$ is measured from the $\hat z$ axis parallel to the crystallographic axis $\langle 111\rangle$. Longitudinal ($\rho$) and Hall resistivity ($\rho_{yx}$) presented here are symmetrized and antisymmetrized, respectively, with respect to the polarity of $H$ and $\theta$. 
We found the demagnetization correction is negligible compared to magnitudes of features reported here (see Supplementary Materials~\cite{SuppInfo}).

\bfig[ht]
\begin{center}
\includegraphics[width=1\linewidth]{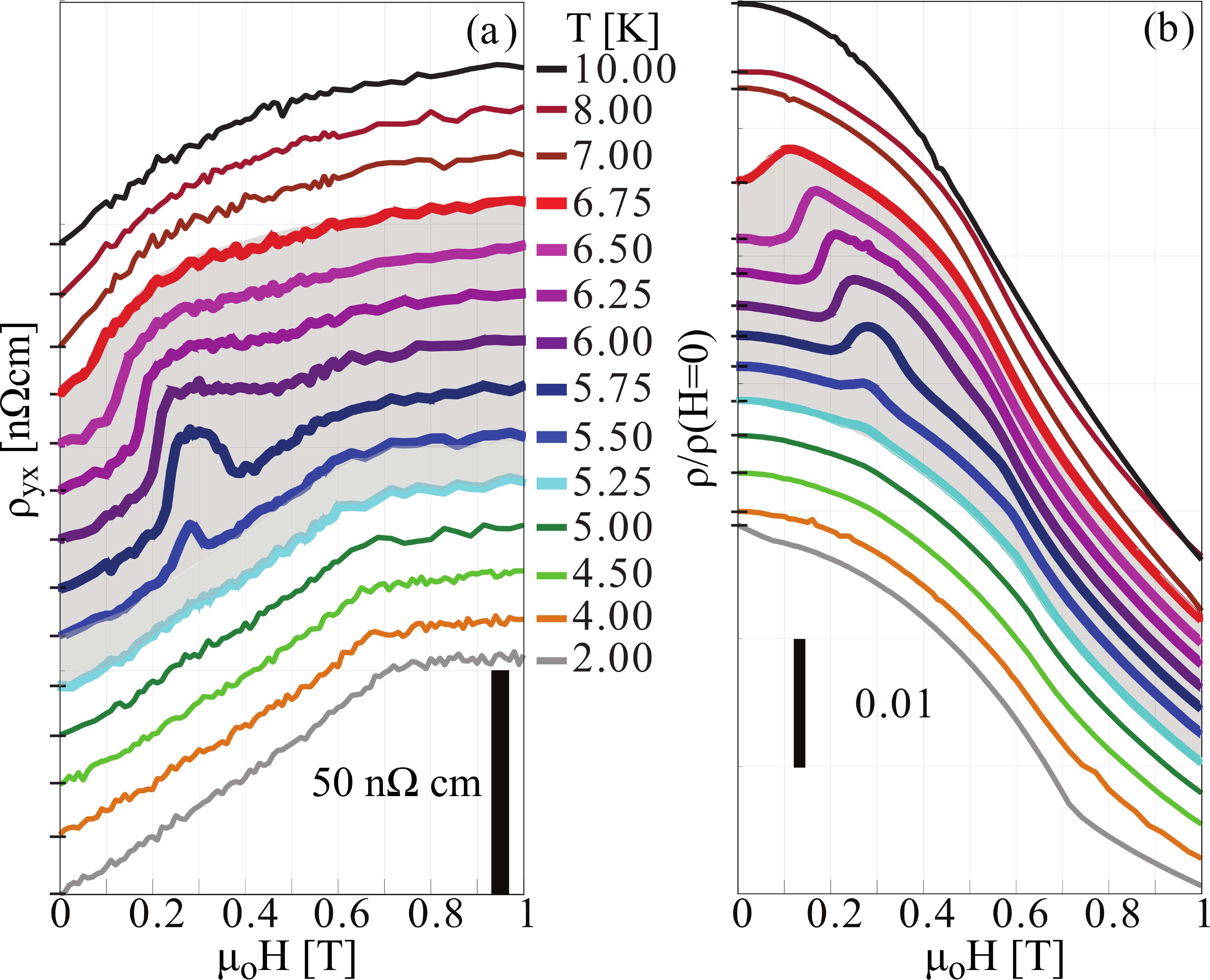}
\caption{\small (a) Hall resistivity $\rho_{yx}(H)$ and (b) fMR $\rho (H)/\rho(H=0)$ as a function of $H$ are plotted with offset for clarity. Gray shaded areas indicates  the $T$-ranges  corresponding to the $A$-phase (thick lines).}
\label{Hsweep}
\end{center}
\efig

{\emph {Results }} 
Unlike epitaxially grown films of MnSi or Mn$_{1-x}$Fe$_x$Si~\cite{Li2013,Yokouchi2014}, the THE signal can be easily identified from the unique $H$-profile  in bulk crystalline samples.  Fig.~\ref{Hsweep}(a) shows  the progression of $\rho_{yx}(H)$ in the $A$-phase (the  shaded area). 
At $T=5.50$ K, $\rho_{yx}(H)$ displays the onset of the THE, persisting throughout the range of $T$ of the shaded region and the narrow bounds of the applied field. As $T$ increases, the small bump in the $H$-profile, starting at around $H\simeq$ 0.3 T at as low as $T= 5.25$ K, develops into a ``peak-like" anomaly and then ``shoulder-like", where $\rho_{yx}(H)$ remains constant until merging to the background. This region of $T$ and $H$ is referred to as the $A$-phase. 
We note the signs of the anomalous Hall signal and THE of \mnfesi~here are found to be opposite from those of pure MnSi, attributed to sufficiently large Fe doping that alters the electronics structure responsible for the sign of the Berry phase contribution and the polarity of the effective carrier~\cite{Chapman2013}.

Similar progression of the $H$ dependence is also found in fractional MR (fMR), defined as $\frac{\rho(H)}{\rho(H=0)}$, shown in Fig.~\ref{Hsweep}(b) in the shaded area.
Entering the $A$-phase with increasing $H$,  MR becomes enhanced  due to the  increased spin scattering arising from the SkL texture simultaneously to the emergence of the  THE signal.  This enhancement collapses back as exiting the $A$-phase, resulting in the peak-like feature in MR as well. 
At higher  $T$'s (6.00 and 6.25 K), MR undergoes similar changes to the Hall signal, exhibiting the shoulder-like $H$-dependence when approaching the spin polarization field ($H_P$). Beyond $H_P$,  it decreases rapidly due to the significantly reduced spin scattering in the polarized state.  . 

\begin{figure*}[ht]
\begin{center}
\includegraphics [width=1\linewidth] {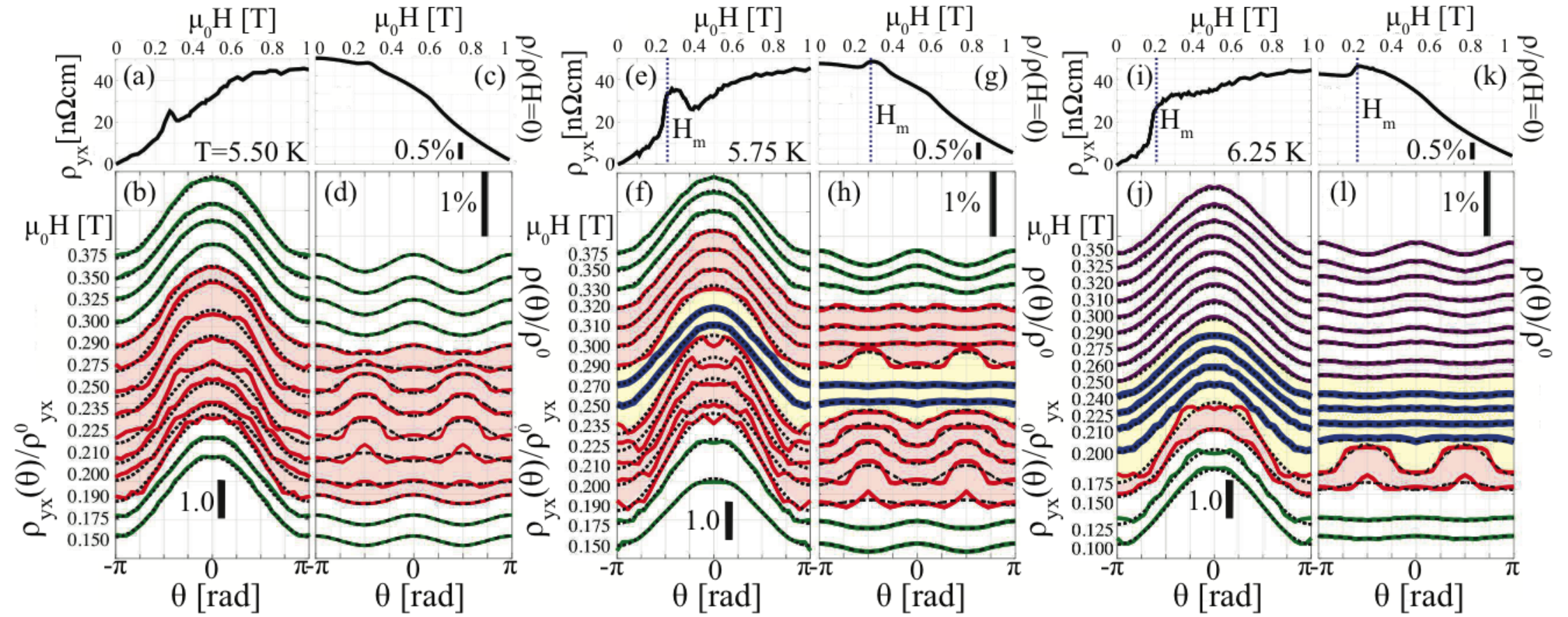}
\caption{\small (a-d) $\rho_{yx}(H)$, $\rho_{yx} (\theta)/\rho_{yx}^0$, $\rho(H)/\rho(H=0)$ and $\rho(\theta)/\rho^0$ at $T = 5.50$ K,
(e-h)  $T=5.75$ K,  (i-j) $T=6.25$ K.  Note both (h) and (l) show the angle-independent region that separates two different $\theta$-dependence. 
The angle sweep curves are shown with offsets for clarity. The regions outside of the $A$-phase are plotted in green. Region I, II and III are marked with red, yellow, and purple shading respectively. 
Broken black lines show $\cos\theta$ and $\cos2\theta$ dependences for Hall and MR, respectively.
}
\label{AllAng}
\end{center}
\end{figure*}

We first establish the baseline of the field orientation dependence outside of the $A$-phase, where the MR and Hall signals unequivocally exhibit  $\cos2\theta$ and $\cos\theta$ dependences, respectively. 
 Fig.~\ref{AngDep} (b, c) displays the angular dependence of the normalized Hall resistivity, $\rho_{yx}(\theta) /\rho_{yx}^0$, where $\rho_{yx}^0 = \rho_{yx}(\theta=0)$ at the given $H$'s and $T$'s of which values lie well outside of the $A$-phase. 
 Since $\rho_{yx}$ in this region is largely contributed from anomalous Hall effect due to the background magnetization ($M$) the $\cos\theta$ dependence indicates smooth rotation of $M$ along $H$. 
 Fig.~\ref{AngDep} (d,e) displays the angle dependence of normalized MR, $\rho(\theta)/\rho^0$, where $\rho^0 = \rho(\theta= 0)$. Two MR curves track the $\cos2\theta$ dependence, which is expected for soft ferromagnetic metals, known as anisotropic magnetoresistance~\cite{McGuire1975}.  Note that the angle sweep at $H =0.15 $ T at 4.5 K shows a distinct curvature near $\theta=0$, which clearly differs from that of 6 K. This  is supposed to arise from  the depinning of the helical propagation direction \cite{Lobanova2016}.
 
Next, we present the results on angular dependences of $\rho_{yx}$ and MR 
at three representative $T$'s  -- 5.50, 5.75 and 6.26 K  -- within the $A$-phase.
Fig.~\ref{AllAng}(a-d) shows $\rho_{yx}(H)$ [(a)], fMR $(H)$ [(c)] and $\theta$-dependences  of the normalized resistivities $\rho_{yx}(\theta) /\rho_{yx}^0$ [(b)] and $\rho_{yx}(\theta) /\rho_{yx}^0$ [(d)] at $T= 5.50$ K, where 
both the THE and enhanced MR begin to emerge. 
In the $H$ range of $0.175 <\mu_0H< 0.250$ T, the angle dependences exhibit strong deviations from the cosine functions, shaded  in (b) and (d). Weak THE signal indicates either low density and/or an imperfect lattice of skyrmions. Similarly, deviations from $\cos2\theta$ observed in MR, \ref{AllAng}(d), point to the onset of the $A$-phase.

Fig.~\ref{AllAng}(e-h) shows the set of data at $T= 5.75$ K. 
Here, both Hall and MR angle data  start with following usual $\cos \theta$ and $\cos2\theta$  at low field [green curves in (f) and (h)]. Then, MR signal  consequently go through two $H$-region of strong deviation from $\cos 2\theta$ function [red shade in (h)], which are separated by the striking $\theta$-$independent$ MR [thick lines  in yellow shade]. Meanwhile, the Hall data also shows the deviations from $\cos\theta$ in two red shaded regions but it recovers the strict $\cos\theta$ form in the yellow region.  
Further increasing $H$, exiting the $A$-phase and entering the conically-ordered state, the cosine function dependences are recovered. 

We note that the narrow  field range of the isotropic MR in the yellow shade (\ie $\rho(\theta)/\rho^0$ remains constant) include the $H_m$ [ Fig. 3(e)], indicating  the SkL lattice is fully formed  and the  magnitude of the resulting gauge field is at its largest. The simultaneous recovery of $\cos\theta$  in Hall  evidences that the SkL plane rotates  freely along the applied field direction, decoupled from the lattice or pinning centers. This  makes it plausible  to have a constant spin-scattering rate that leads to constant MR upon rotating  the orientation of $H$.  

Fig.~\ref{AllAng}(i-l) shows the same set for $T=6.25$ K. Here,   $\rho_{yx}(\theta)/\rho^0$  shows a clear deviation from $\cos\theta$ only in the low $H$ [red shade in (j)] but it remains  in the $\cos\theta$  form for the rest of the $H$'s. On the other hand, MR exhibits  the same  $\theta$-independent region [yellow shade in (l)] followed by the strong deviation from $\cos2\theta$ form [red shade]. In high $H$, it smoothly transforms to  $\cos2\theta$ form, which is different from what was observed at lower $T$ [Fig.~\ref{AllAng}(f) and (h)].

Based on these observations, we classify the $A$-phase and the immediate surroundings into three separate regions in the $H$-$T$ phase diagram:

\noindent(1) Region I : Red-shaded regions [Fig.~\ref{AllAng}(b,d,f,h) and (j,l)]  are marked by  the deviations from $\cos\theta$  (Hall) and $\cos2\theta$ (MR) dependence.  The  irregular $\theta$-dependence in this region  is ascribed to crystallogrphic  orientation dependence or pinning of skyrmions during the nucleation and annihilation of individual skyrmions  or the SkL. 
In low $T$ of this region [Fig. 3(b,d) ],  the SkL is still premature so that the resulting gauge field responsible for THE must be weak. Thus nucleation of skyrmions and the SkL encounter pinning from the impurities and the crystalline anisotropy~\cite{Neubauer2009, Bauer2010, Bauer2012}, preventing it from closely following the applied field direction resulting in strong deviations from $\cos\theta$ dependence.  At  higher $T$,  this region occurs at at nucleation (lower $H$) and annihilation (higher $H$), from which we deduce that the effect of crystalline anisotropy and pinning effect is most prominent as entering and exiting the $A$-phase. 

\noindent(2) Region II : Yellow-shaded regions [Fig.~\ref{AllAng}(f,h) and (j,l)] are marked by the constant MR upon rotating $H$, indicating the emergence of the fully developed SkL decoupled completely from the crystalline lattice and pinning sites. 
Here the Hall signal reaches its maximum, simultaneously  recovering $\cos\theta$ dependence, from which we infer that the SkL is fully formed and generating the maximum magnitude of the gauge field. 
 This region occupies a narrow range of $H$ at the center of the $A$ phase, although it extends close to $T_C$.

\noindent(3) Region III : Purple-shaded region [Fig.~\ref{AllAng}(j,l)] is marked by  the smooth recovery  $\cos2\theta$  after $\theta$-independent MR. We note, in this region, the $H$ dependences of THE and MR exhibit ``shoulder-like" features, where  the magnetic state in the $H$ range of constant MR and Hall should be discernible from the conical state.
Approaching $T_C$ near the upper $T$ and $H$ boundary of the $A$-phase, the SkL maintains the long range translational symmetry while the individual skyrmions are polarized and thus,   the emergent field  is reduced to to zero smoothly. This results in 
$\cos2\theta$ dependence being recovered in the MR, with no sign of exiting the $A$-phase (\ie without having Region I in the high $H$).  It  is  consistent with the shoulder-like $H$-dependence  in $\rho_{yx}$.

\begin{figure}[htb]
\begin{center}
\includegraphics[width=1.0\linewidth]{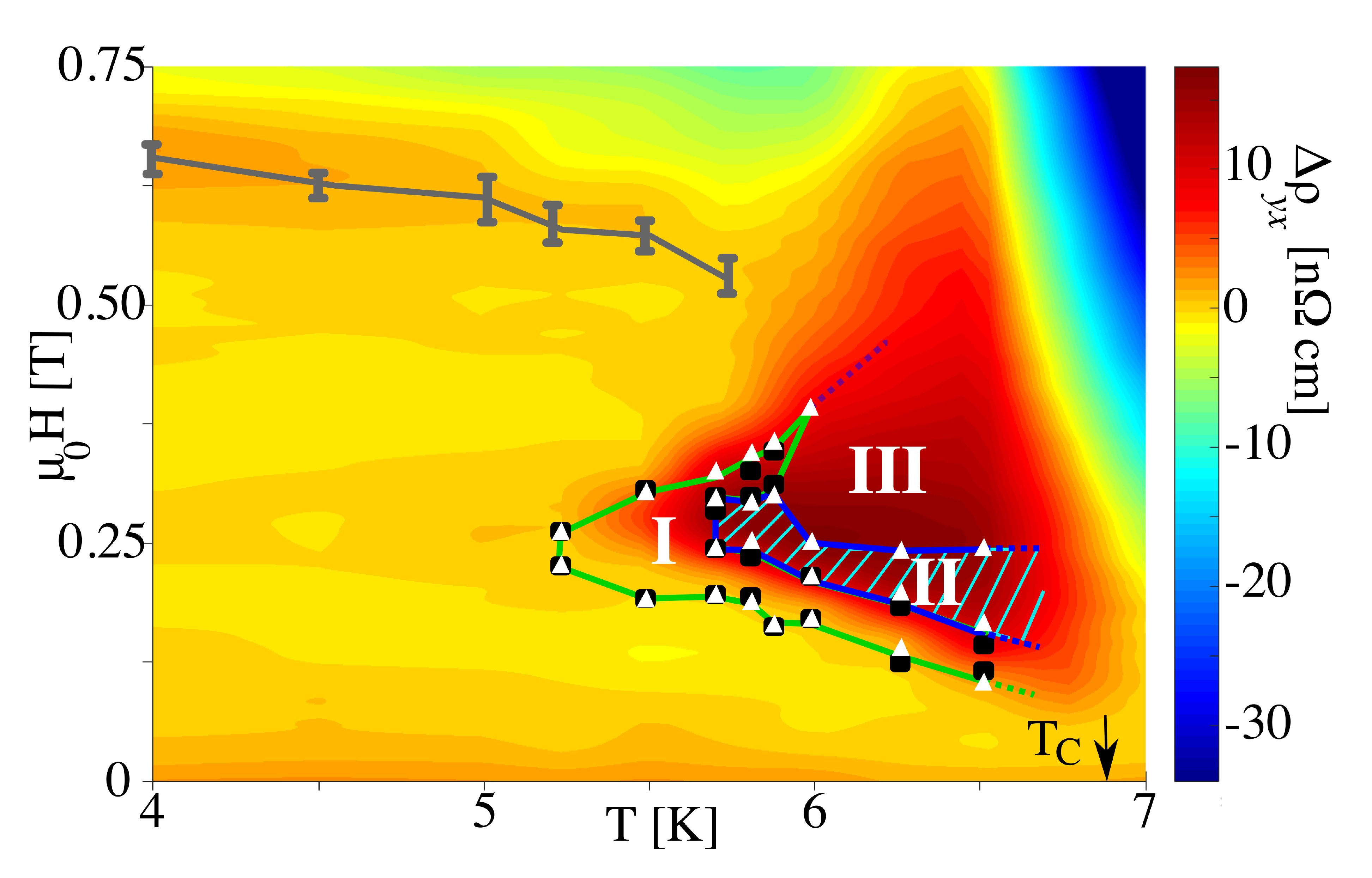}
\caption {\small Color map generated from $\Delta\rho_{yx}$, calculated by subtracting the linear background below spin polarization field $\rho_{yx} (H)$.
The gray line refers to the onset of spin-polarization marked from the kink in $\rho_{yx}(H)$ shown in Fig.~\ref{Hsweep}(a). Three distinct Regions I, II and III are explicitly identified by the field orientation and magnitude dependence (See text).
}
\label{Surf}
\end{center}
\end{figure}

The resulting phase diagram is presented in Fig.~\ref{Surf}, where Region I, II and III are indicated. The colormap is generated from $\Delta \rho_{yx} (H)$, obtained by subtracting the $T$-dependent linear background up to $ H=H_p$ from $\rho_{yx}(H)$ [Fig. ~\ref{Hsweep}(a)]. The overlaid symbols correspond to  the values of $(T,H)$ of $\rho(\theta)/\rho^0$ (white triangle) and $\rho_{yx}(\theta)/\rho_{yx}^0$ (black square), from the different angle dependences in the Hall and MR data.  These clearly illustrate the three different regions of the $A$-phase discussed above. 
We examined a different Fe doping (8\%) sample and found the  angle and field dependence are consistent ~\cite{SuppInfo}. This  supports our conclusion that these three regions are not specific to a particular Fe contents but can be a general trait for the $A$-phase.

{\emph {Discussion }}
It is intriguing to juxtapose our classification of the $A$-phase to what was found in the LTEM images in pure MnSi crystals: Multi-domain SkL, single-domain SkL and skyrmion glass~\cite{Yu2015}. 
The multi-domain SkL consists of separate hexagonal skyrmion lattices existing with distinct boundaries between different orientations similar to grain boundaries of polycrystalline materials. In the LTEM observation in Ref.~\cite{Yu2015}, this state appears at low $H$ area of the $A$-phase, similar to our Region I. 
The multi-domain SkL implies long range translational symmetry has not been fully established yet. The domain boundary is likely to form along the impurities, in analogy to the superconducting vortex. Deformation from the regular triangular lattice would result in a weaker THE signal, yet still yield the peak-like $H$ profile. The increased spin scattering due to the nucleating skyrmions also leads to the increase in MR in this narrow range of field. 

 On the other hand, the fully formed single-domain SkL phase of Region II generates the largest and most robust magnitude of the THE, while it is oblivious to the presence of disorder, and the planes of the SkL, and thus the orientation of the gauge field, closely follow the direction of $H$.

The state in further increased $H$ is denoted as the skyrmion glass phase ~\cite{Yu2015}. Here the individual skyrmions are inferred to be in the process of being polarized, as if dissolving into the ferromagnetic state, while the triangular SkL is still maintained. This corresponds to Region III, where the spins comprising the SkL are diffused by applied $H$, yet the periodic structure of the SkL maintains a non-zero gauge field. It seems that the transition from single domain SkL to skyrmion glass is smooth and adiabatic. 
Such change is known to have little effect on the THE~\cite{Jalil2014}, which is consistent with $\cos\theta$ dependence of $\rho_{yx}$ in our data.

Overall, we find the role of impurities as pinning centers is feeble and ineffective. 
In fact, there is no clear argument nor evidence to distinguish the effects originated from  of weak crystalline anisotropy and originated from pinning centers, as both would leave distinct  non-trivial angular dependences  within the $A$-phase in the electrical transport data. This is consistent with the theoretical consideration that strong interaction with impurity sites occurs at the time of nucleation~\cite{Choi2016}.  
 From the way Region I surrounds Region II [Fig.~\ref{Surf}]  we deduce that  in Region I, the nucleation of the skyrmions and the SkL is the most receptive to the crystalline anisotropy and the presence of quenched disorders.
 
Region II resembles the previously reported $A$-phase core~\cite{Lobanova2016} in pure MnSi in terms of constant $\rho(\theta)$ upon rotating field. However, robust $\cos\theta$  dependence indicates that the emergent field freely rotates along with the applied field, independent from underlying crystallinity and pinnings.
In the comparison with pure MnSi, the size of the magnetic moment is reduced by a factor of two and $T_C$ by a factor of  four. Also,  the residual resistivity increases 20 times in \mnfesi~which leaves the mean free path of electron only up to a few percent of the lateral size of the SkL~\cite{Grigoriev2009, Chapman2013}. 
Despite these effects, the remarkable robustness found in electrical transport properties  in the $A$-phase is expected to play an essential role in developing skyrmion-based devices and hosting materials and is particularly noticeable in Region II and III.

{\emph{Summary }} We identify three distinct regions within the $A$-phase in \mnfesi, reflecting a series of distinct SkL states from nucleating out of conical ordering to vanishing into spin polarized state as a function of $T$ and $H$. 
 The roles of crystalline anisotropy and pinning due to Fe impurities is only discernible in Region I. Experimental distinction of the multiple magnetic states provides the basis for flexible geometries in devices utilizing skyrmion-based electrical responses. 
 We looked for possible signatures that distinguish Region I and II in magnetization measurements for a few different angles but found such features are within the noise-level  and far from conclusive \cite{SuppInfo}. 
 Further investigation will definitely be beneficial to understanding the nature of these states.


\vspace{0.1in}
\noindent{\bf {Acknowledgment}} This work was supported by the US DOE, Basic Energy Sciences, Materials Sciences and Engineering Division under Award Number DE-SC0006888. 

\vspace{0.1in}
\noindent $^{\dagger}$Current Address: Department of Applied Physics, Stanford University, Palo Alto CA 







\onecolumngrid
\newpage

\setlength{\tabcolsep}{6pt} 
\renewcommand{\arraystretch}{1.25} 

\renewcommand*{\citenumfont}[1]{S#1}
\renewcommand*{\bibnumfmt}[1]{[S#1]}

\setcounter{figure}{0}
\renewcommand{\thefigure}{S\arabic{figure}}

\setcounter{equation}{0}
\renewcommand{\theequation}{S\arabic{equation}}

\setcounter{table}{0}
\renewcommand{\thetable}{S\arabic{table}}

\setcounter{page}{1}

\renewcommand{\theequation}{S\arabic{equation}}
\renewcommand{\thefigure}{S\arabic{figure}}

\section*{\Large{Supplemental Material}}

\centerline{\bf {Multiple magnetic states within the $A$-phase determined by field-orientation dependence of \mnfesi }}

\vskip2mm

\centerline{Peter E. Siegfried, Alexander C. Bornstein, Andrew C. Treglia, Thomas Wolf, Minhyea Lee} 

\vskip5mm

\twocolumngrid
\subsection{S1. Sample Preparation and measurements}
Single crystals of \mnfesi~  used were grown by the Bridgman technique  and cut with typical dimensions $\approx$ 2.7 $\times$0.8 $\times$0.1 mm$^3$. $T_C  = 6.9 $ K is determined from the inflection point of the $T$ dependence of resistivity. Six contacts of gold wire and silver paint were made  with a contact resistance $\le 1-2$ $\Omega$.   The Hall and MR signals were measured simultaneously~\cite{Suslov2010}, using 4 -- 10 mA  of applied current, which is  at least  an order of magnitude lower than the depinning current for MnSi~\cite{Schulz2012}.

\subsection{S2. Demagnetization Correction  in the  Angle Sweeps of the MR}


\bfig[h]
\begin{center}
\includegraphics[width=0.9\linewidth]{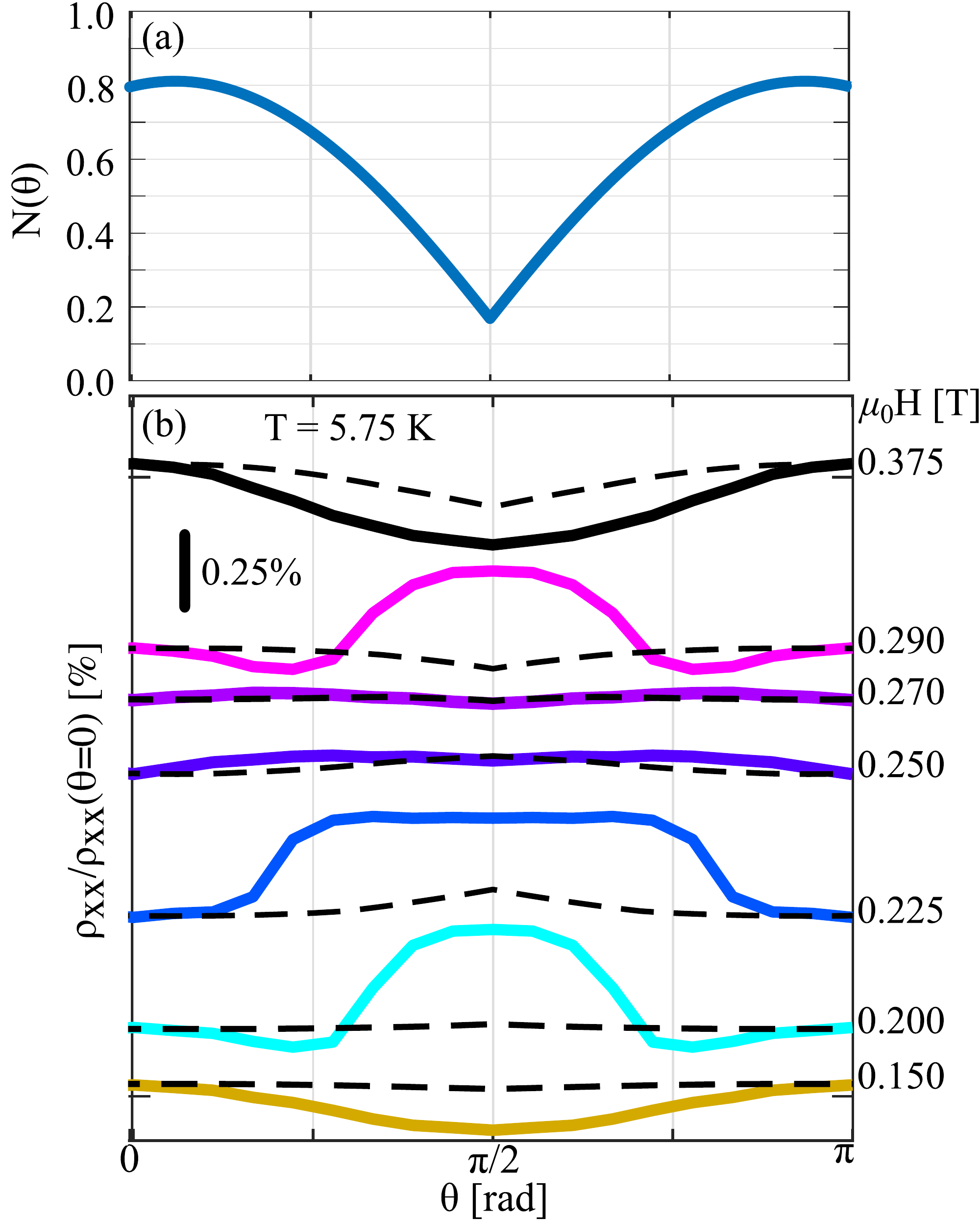}
\caption {(a) Demagnetization correction factor $N$ as a function of angle $\theta$ for samples with a rectangular prism geometry. (b) Comparison of the angular dependence of MR data (solid lines) and estimation of  the MR variation from the demagnetization correction (dotted line). Both the functional form and the magnitude are clearly discernible  between the data and correction factor ;  demagnetization alone cannot account for the features shown in the angular dependences.  }
\label{demag}
\end{center}
\efig

Contribution to  demagnetizing fields is tracked down  as the applied field is rotated, which  arises from a  rectangular prism shape of the  samples used in our  magnetotransport measurements. Geometry of measurement is shown is  in Fig. 2 (a) in the main text.
We  calculate  the demagnetization correction factor, $N(\theta)$, \cite{Osborn1945, Aharoni1998} using the sample dimensions. Fig. S1 (a) displays  $N(\theta)$, of which 
maximum, 0.8,  occurs near $\theta=0^{\circ}$, i.e. the field angle lies out of plane of the face of the largest area and the minimum value, 0.16, at $\theta=90^{\circ}$. 
Next, using  the magnetization vs applied field   at $\theta =0$ and $\theta = 90^{\circ}$ measured on the same sample (data not shown), we deduce the variation of  the magnitude of field as a function of $\theta$ and the corresponding MR deviation. 

In Fig. ~\ref{demag} (b),  the dotted line shows  $\rho_{xx}(\theta)/ \rho_{xx}(\theta=0)$ the contribution of MR variation arising solely from $N(\theta)$.  Characteristics of measured data displayed in solid lines are unambiguously  discernible from the demagnetization correction in both  magnitude of variation and  functional  form with respect to $\theta$.  
As  $H$ increases, the demagnetization correction reaches close to  $50\%$ as shown  at  $\mu_0H = 0.375 $ T.  This can affect the angular dependence in transport quantity. 

However,  this range of field, i.e.  $H>0.3$ T corresponds to Region III of the $A$-phase or spin polarized state. The boundary between Region II and III  lies at lower fields as dictated from the $H$-dependence, the demagnetization corrections do not  affect our observation.
 In fact, the size of error bars is at most  the same or less than  the size of symbols (white triangles and black squares) for data points marking the boundaries of Region I, II and III in Fig. 4 of the main text.   We find the qualitatively important features of the phase diagram within the $A$-phase  remain robust in the presence of demagnetization fields, which is also consistent with what was reported previously \cite{Bauer2012}.

\subsection{S3. Comparison of $H$-dependence  of MR for 8\% and 10\% Fe doped samples }

\bfig[h]
\begin{center}
\includegraphics[width=0.95\linewidth]{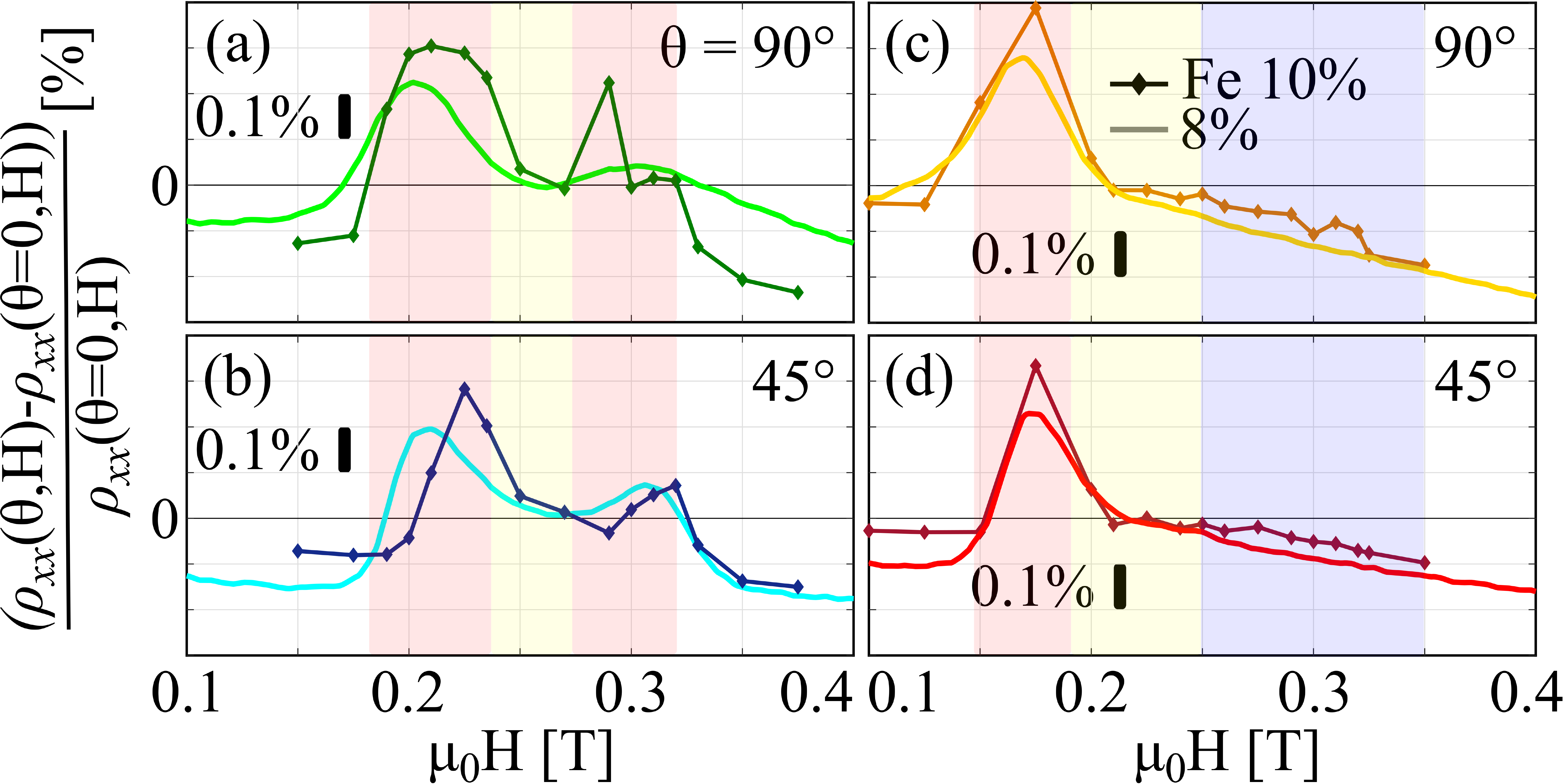}
\caption { $H$ dependence of normalized  MR with respect to $\theta =0^{\circ}$, i.e.  $\frac{\rho_{xx}(\theta)-\rho_{xx}(\theta =0)}{\rho_{xx}(\theta =0)}$ for two Fe doping contents, 8\% (line) and 10\% (symbol+line) at $\theta = 45$(b,d) and $90^{\circ}$ (a,c). $H$-sweep in  (a) and (b) are  at $T= 5.75$ for 10\% sample 
and 6.50 K 
 for 8\%, where both samples shows the peak-like  behavior in MR and Hall effect.  (c) and (d)  are at $T= 6.25$ (10\%) and 
7.0 K (8\%), where shoulder-like H-dependence is exhibited.   Red, orange and purple shades  correspond to Region I, II and III respectively. }
\label{figs3}
\end{center}
\efig

Fig.~\ref{figs3} (a) and (b)  shows the  $H$-dependence  of the MR signal in two different Fe doped samples -- 10\% and 8\%  at two angles of $\theta = 45$ and 90$^\circ$. The data in (a) and (b) are taken at $5.75$ K for the 10\% sample and at 
6.5 K for the 8\% sample, \black the temperatures where each sample exhibits the ``peak-like" $H$-dependence in MR and Hall in the $A$-phase.  The feature occurring  at different $T$'s is due to the difference in  $T_C$ (6.9 K  for 10\% vs 8.0 K for 8\%). Similarly,  the shoulder-like features of MR of the two samples are shown in (c) and (d), which are taken at $T=6.25$ K and 7.0 K for 10\% and 8\% sample respectively.  Both sets of data (a,b) and (c,d) correspond to the vertical cuts ( i.e. constant temperature)  of the phase diagram in the Fig. 4 in the main text and   the colors of shade indicates different Regions within the $A$-phase. Identical features of the maxima and valleys in the $H$ dependence
  occur at the same  Regions as well as outside of the $A$-phase.  Thus, the observed magnetotransport characteristics are  general properties of  the $A$-phase,  rather than traits  of individual sample or specific disorder configuration.

\subsection{S4. Magnetization at the vicinity and inside of  the $A$-phase }

DC magnetization measurements were performed on 10\% Fe doped sample ( the same piece that used in the electrical measurement)  as a function of applied field, using a commercial SQUID magnetometer (MPMS, Quantum Design).   We particularly focused on the vicinity and inside of the $A$-phase with  three different  directions ($0, 45$ and $90^{\circ}$) of  applied field, as shown in Fig. ~\ref{figs4}. 
Arrows  indicates  features that may be possibly associated with the  onset of $A$-phase and different Regions discussed in the main text. However the  magnitudes of those features are very feeble and remains at the noise level. Moreover, these features persist  at $   T= 8$ K $>T_C = 6.9 $ K, where  the characteristics related to the $A$phase in the electrical measurement disappear. This remains a puzzle. 

In order to  properly identify the features for Region I, II and III in the magnetization data, the  full angle dependence of 
magnetization measurement  would be  much desirable, which is  beyond the scope of our current  study.

\bfig[h]
\begin{center}
\includegraphics[width=0.8\linewidth]{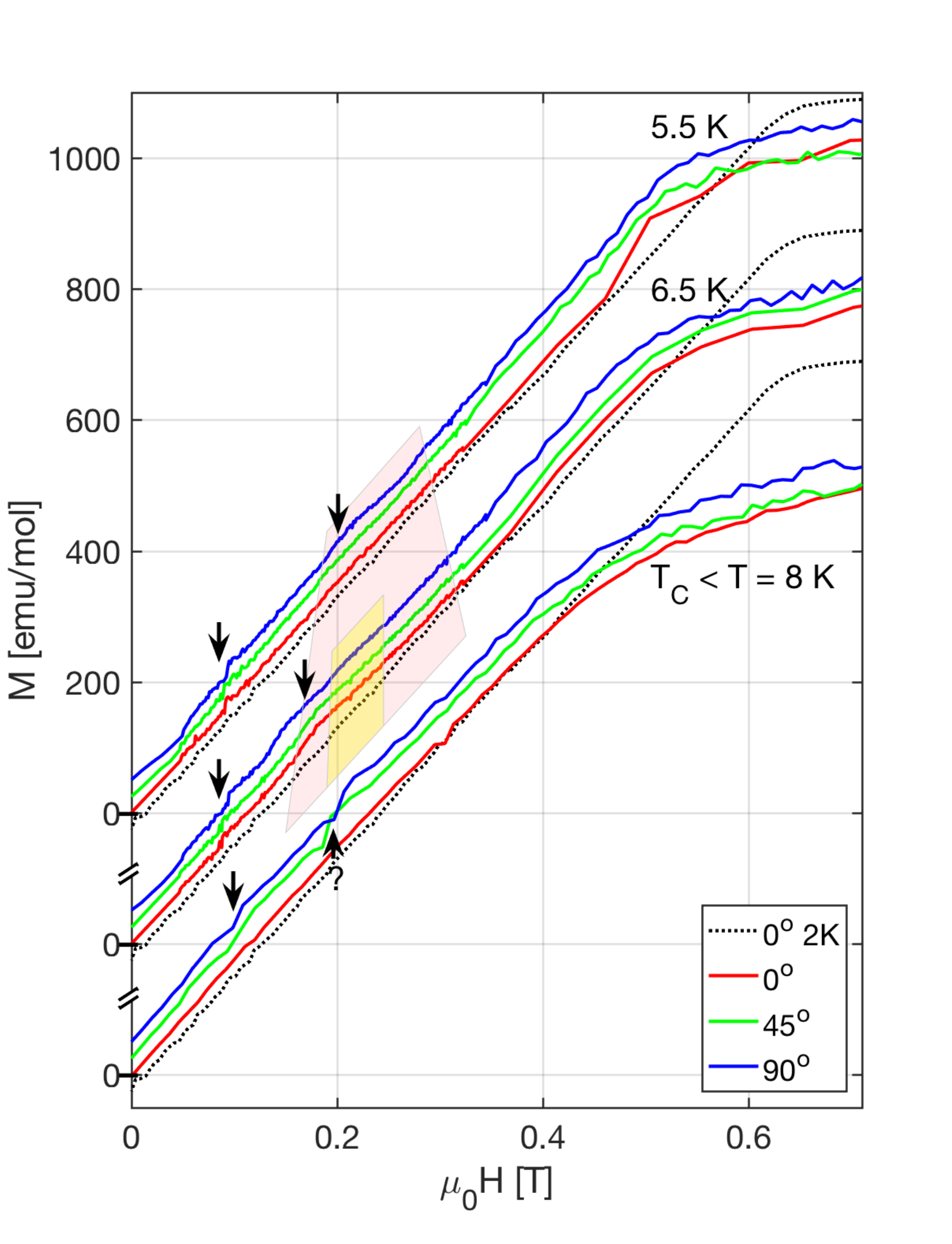}
\caption { $H$ dependence  the magnetization ($M$)  in the vicinity and inside of the $A$-phase.  All curves are offset for clarity and demagnetization factor corrected. Broken line corresponding to $M(H)$ at $T=2$ K are shown for a comparison.  Arrows indicate the features that may be possibly  associated with the $A$-phase. However, their magnitudes are at the level of noise for the measurement.  Red and yellow shading indicate the $A$-phase and the region of isotropic MR, respectively, both of which are identified in the the angular dependence  of electrical transport data.}
\label{figs4}
\end{center}
\efig

\end{document}